\documentclass[12pt]{amsart}
\usepackage{amsfonts,amsthm}
\usepackage{amssymb,amsmath}
\usepackage{amsfonts,mathrsfs}
\usepackage{xcolor}

\begin{document}

\noindent{\bf Felix Klein's ``Uber die Integralform der Erhaltungss\"atze und
die Theorie der r\"aumlich-geschlossenen Welt'': an English translation}

\bigskip
\centerline{Chiang-Mei Chen$^{1,2}$, James M. Nester$^{1,3,4}$ and Walter Vogel$^5$}

\medskip

\noindent $^1$ Department of Physics, National Central University, Chungli 32001, Taiwan

\noindent $^2$ Center for High Energy and High Field Physics (CHiP), National Central University,
Chungli 32001, Taiwan

\noindent $^3$ Graduate Institute of Astronomy, National Central University, Chungli 32001, Taiwan

\noindent $^4$ Leung Center for Cosmology and Particle Astrophysics, National Taiwan University, Taipei 10617, Taiwan

\noindent $^5$ Department of Chemistry, National Central University, Chungli, 32001, Taiwan,

\medskip
\noindent {\small email: cmchen@phy.ncu.edu.tw, nester@phy.ncu.edu.tw, vogelw@ncu.edu.tw}

\bigskip

\noindent{\textbf{Abstract}}

\emph{We present an English translation of a third 1918 paper by Felix Klein which follows up on his earlier work.}

\section{Translator's Preface}

In 1918, following up on his paper~\cite{KleinI} about David Hilbert's Foundation's of Physics I~\cite{Hilbert} (which led to Noether's famous work~\cite{Noether, KS}), Felix Klein published another work~\cite{KleinII} about the law of conservation of energy and momentum in Einstein's theory of gravitation, general relativity (for English translations of these two papers see~\cite{KleinIeng},~\cite{KleinIIeng}). The second work, which has for too long been neglected, includes some interesting analysis regarding the gravitational energy-momentum  expressions of Einstein,  Hilbert, Lorentz and Weyl. For the detailed story concerning the related exchanges between Einstein, Hilbert and Klein and the inception of Noether's theorems see Refs.~\cite{Brading05, KS, Rowe99}.
Klein followed these two works with a third related paper~\cite{KleinIII}.
Regarding this paper Einstein remarked: ``Dear Colleague, I am thrilled with your new paper~\cite{KleinIII} like a child who gets a piece of chocolate from his mother.''~and ``Your paper appealed to me very much.''~\cite{cpae9}.

Some years ago the senior member of our team (JMN) began, relying on his long unused undergraduate German and Google Translate, to make a translation of Klein's papers, which we believe include some long forgotten insights. We are fortunate to have recently acquired the help of a native German speaker (WV) to refine our effort into a presentable form. We feel that our translation has now finally reached a form where it can be useful to others, and so want to share it with anyone who may be interested.

The page by page layout, the equation numbers, and footnotes in this version of our translation are from the paper as it appears in Vol.~1 of Klein's collected works~\cite{KleinII}---so anyone who cares to can easily compare our translation with the original. We chose to follow the Klein collected works version, as it includes some additional footnotes that do not appear in the journal version; readers may find the remark regarding Emmy Noether especially interesting. Our translation is a work in progress. We welcome corrections and comments on the translation and on any errors.

\newpage

\textbf{XXXIII. On the Integral Form of conservation Laws and the Theory of the Spatially Closed World.}\\

\footnotesize{[News of the Kgl.\ Society of Sciences at G\"ottingen. Mathematical-physical class. (1918). Submitted at the meeting of 6 December 1918.\footnote{Submitted for printing at the end of January 1919.}]}
\bigskip

------------------------
\bigskip

In my note of July 19, 1918, I tried to gain an overview of the various forms which one can give in Einstein's gravitational theory
for the differential laws for the conservation of momentum and energy; my task today is in the first place to comment on the integral form of the conservation laws which Einstein set up as his preferred form of the differential laws.
In connection with this I will treat Einstein's theory of the spatial-closed world and the modification, which has been found by de Sitter.\footnote{
The relevant publications are:

Einstein.
\begin{itemize}
\item[1.] Cosmological Considerations on the General Theory of Relativity.
Meeting reports of the Berlin Academy of February 8, 1917.
\item[2.] Criticism of a Solution Given by M. de Sitter
to the Gravitational Equations, ibid, March 7, 1918.
\item[3.] The Energy Law in the General Theory of Relativity, ibid,
16 May 1918.
\end{itemize}

de Sitter.
\begin{itemize}
\item[] In various notices published by the Amsterdamer Akademie, 1917, as well as in a comprehensive article series in
the Monthly Notices of the R. Astronomical Society: On Einstein's theory of gravitation and its astronomical consequences
(see in particular the Final Part III of November 1917).
\end{itemize}}
The physical questions are barely touched, the aim is to clarify the mathematical connections completely; I feel a certain satisfaction that my old ideas of 1871--72 are of decisive importance.\footnote{
See in particular:
\begin{itemize}
\item[1.] On the so-called non-Euclidean geometry.\ Math.\ Annalen (1871), Bd. 4. [Ab. XVI of this issue.]
\item[2.] The Inaugural Program: Comparative Reflections on Recent Geometric Researches, Erlangen, 1872. [Ab. XXVII of this edition.]
\end{itemize}}
To what extent progress has been made, the reader may decide by a comparison with the representations of  other authors.

\newpage

\hfill XXXIII. Integral Form of Conservation Laws, etc. \hfill 587
\bigskip

I will first recall the following results: The conservation laws in
the form which I named after Lorentz (formula (42) of the preceding note):
\begin{equation}%1
\frac{\partial \left( \frak{T}^\sigma_\tau + \frac{1}{\chi} \frak{U}^\sigma_\tau \right)}{\partial w^\sigma} = 0.
\end{equation}
We write
\begin{equation}%2
\frac1{\chi} \frak{U}^{*\sigma}_\tau = \frak{t}^\sigma_\tau,
\end{equation}
we obtain the Einstein form of the conservation law:
\begin{equation}%3
\frac{\partial \left( \frak{T}^\sigma_\tau + \frak{t}^\sigma_\tau \right)}{\partial w^\sigma} = 0
\end{equation}
(Formula (44) of the previous note).\footnote{
[This whole paragraph could be significantly shortened after the
first publication at this point, sign changes which have been necessary in the previous note, have already been taken into account in reprinting in this edition. Mr Vermiel had drawn my attention to the necessity of these changes of sign, and he has thankfully supported me in many of the calculations needed for the following considerations. K.]}

Now it will correspond to Einstein's basic assumption, if I further for
$$ \hbox{the} \;\; \frac{\frak{U}^\sigma_\tau}{\chi}, \quad \hbox{or the} \;\; \frac{\frak{U}^{*\sigma}_\tau}{\chi} $$
briefly term as the gravitational components of the energy (caused by the choice of arbitrary co-ordination and the respective approach).
Furthermore, I will describe the following components of the ``total energy'' in abbreviated form with the letter $V$, or $\frak{V}$:
\begin{equation}%4
\frak{T}^\sigma_\tau + \frac1{\chi} \frak{U}^\sigma_\tau = \frak{V}^\sigma_\tau, \qquad \frak{T}^\sigma_\tau + \frac1{\chi} \frak{U}^{*\sigma}_\tau = \frak{V}^{*\sigma}_\tau.
\end{equation}

It is a peculiarity of my following account, to which I am referring in advance, that I have the $\frak{U}$ and $\frak{U}^*$ (or also the $\frak{V}$ and $\frak{V}^*$) --- which both have their advantages --- always side by side; one
then sees more clearly how far in the integral forms of the conservation statements a subjective moment comes into play.

For the convenience of the reader, I set the underlying definition of the corresponding Latin letters according to formulae (16), (55)
of the previous note. One has:
\begin{eqnarray}%5
2 U^\sigma_\tau &=& K \delta^\sigma_\tau - \frac{\partial K}{\partial g^{\mu\nu}_\sigma} g^{\mu\nu}_\tau - \frac{\partial K}{\partial g^{\mu\nu}_{\varrho\sigma}} g^{\mu\nu}_{\varrho\tau} + \frac1{\sqrt{g}} \frac{\partial \left( \frac{\partial \sqrt{g}K}{\partial g^{\mu\nu}_{\varrho\sigma}} \right)}{\partial w^\varrho} g^{\mu\nu}_\tau,
\\
%6
2 U^{*\sigma}_\tau &=& G^* \delta^\sigma_\tau - \frac{\partial G^*}{\partial g^{\mu\nu}_\sigma} g^{\mu\nu}_\tau.
\end{eqnarray}

\newpage

\noindent 588 \hfill On the Erlangen program \hfill
\bigskip

I have in (5), as in all my previous notes
following Hilbert's original notation, included the square root $\sqrt{g}$.
If one wishes to have a complete connection with Einstein's method of identification,
we must take $\sqrt{-g}$ everywhere.
This change has no influence on the final formulas (1) to (3);
it is nevertheless appropriate,
so that the quantities subject to direct observation are always given real components;
it is therefore also to be assumed as valid.

\bigskip
\begin{center}
\textbf{I. The integral theorems for closed systems of the ordinary theory.}

\medskip
\S 1.

\medskip
\textbf{From the vectorial notation of multiple integrals.}

\medskip
\textbf{First introduction of $I_\tau$ or $I_\tau^*$.}
\end{center}
\bigskip

Wherever one has to deal with the transformation of multiple integrals, the usual notation, e.g., $\int \int f(xy) dx dy$, is not appropriate. The increments $dx$, $dy$ are nevertheless to be thought off in different directions, therefore belonging to two different vectors, so that already something is gained, when you write $\int \int f(xy) d'x d''y$. The notation becomes even more clear if the vectors $d'$, $d''$ are not exactly parallel to the two coordinate axes, but arbitrarily chosen and the product $d'x d''y$ accordingly, replaced by the content of the parallelogram enclosed between the two vectors. Thus we come to the notation
\begin{equation}%7
\int\int f(xy) \cdot \left| \begin{array}{cc}
                                   d'x & d'y \\
                                   d''x & d''y
                            \end{array} \right|,
\end{equation}
which I like to call Gra{\ss}manns, because they are the quantities in Gra{\ss}mann's expansion theorem of 1861: the formula is the mobility, which we put into the conception of the multiple integral, which is better adapted for the purpose.  For the purpose of particular evaluation one will of course be able to go back to the ordinary notation at any moment.

For all transformation considerations, however, (7) is preferable.
If we set, e.g, $x = \varphi(\xi, \eta)$, $y = \psi(\xi, \eta)$, then from (7)
it is immediately clear why in the transformation formula of the integral the
Jacobian functional determinant comes in. For one has identically:
$$ f(xy) \cdot \left| \begin{array}{cc}
                             d'x & d'y \\
                             d''x & d''y
                      \end{array} \right| = f(\varphi\psi) \cdot \left| \begin{array}{cc}
                                                                               \varphi_\xi & \varphi_\eta \\
                                                                               \psi_\xi & \psi_\eta
                                                                        \end{array} \right| \cdot \left| \begin{array}{cc}
                                                                                                                d'\xi & d'\eta \\
                                                                                                                d''\xi & d''\eta
                                                                                                          \end{array} \right|. $$
Having said that, we shall now come to certain triple integrals, which are as follows:

\newpage

\hfill XXXIII. Integral form of conservation laws, etc. \hfill 589
\bigskip

\begin{equation}%8
I_\tau = \int \int \int \left|
\begin{array}{ccc}
  \frak{V}^I_\tau & \dots & \frak{V}_\tau^{IV} \\
  d'w^I & \dots & d'w^{IV}  \\
  d''w^I & \dots & d''w^{IV}  \\
  d'''w^I & \dots & d'''w^{IV}
\end{array} \right|,
\end{equation}
or the others, that I call
\begin{equation}%9
I_\tau^*
\end{equation}
and which result from the foregoing, by replacing the $\frak{V}^\sigma_\tau$ with $\frak{V}^{*\sigma}_\tau$.
--- these integrals have to be extended over some part of a ``hypersurface'' located in the four-dimensional world $w^I \dots w^{IV}$; $d'$, $d''$, $d'''$ denote three independent vectors, which each extend in the tangential direction from the individual point of the hypersurface.

From the differential laws (1), (3) to which the $\frak{V}^\sigma_\tau$ resp. $\frak{V}_\tau^{*\sigma}$ obey, one will --- assuming the usual continuity and uniqueness properties for the $\frak{V}$ --- conclude from the start, \textit{that these $I_\tau$, resp.~$I_\tau^*$ vanish if one takes their integration domain closed in such a way that it delimits a certain segment of the world.}
As a matter of fact, the $I_\tau$ transform in the known way into the enclosed part of the world extended quadruple integrals
\begin{equation}%10
I_\tau = \int\!\int\!\int\!\int \left( \frac{\partial \frak{V}^\sigma_\tau}{\partial w^\sigma} \right) \cdot
\left| \begin{array}{ccc}
  dw^I & \dots & dw^{IV} \\
  d'w^I & \dots & . \\
  d''w^I & \dots & . \\
  d'''w^I & \dots & .
\end{array} \right|,
\end{equation}
and similarly the $I_\tau^*$, where the integrands themselves vanish easily because of the conservation laws (1), (3). ---

Our particular interest, however, is focused on how the $I_\tau$, $I_\tau^*$ for \emph{affine} transformations of the $w$ behave, if one thus submits the $w$ to linear transformations with constant coefficients:
\begin{equation}%11
{\bar w}^\varrho = a^\varrho_1 w^I + \cdots + a^\varrho_4 w^{IV} + c^\varrho.
\end{equation}
Our vectorial notation has now proven itself here. We know
from the developments of the previous note, that the $V_\tau^\sigma$ or $V_\tau^{*\sigma}$ with
the transformations (11) behave as mixed tensors; from them
the $\frak{V}_\tau^\sigma$ or $\frak{V}_\tau^{*\sigma}$ arise through multiplication by $\sqrt{g}$ (or $\sqrt{-g}$).
After that it is obvious that the integrands $dI_\tau$ or $dI_\tau^*$
transform as ``contragredient'' vectors.
This means that they undergo the homogeneous linear substitutions derived from (11):
$$ dI_\tau = a_\tau^I \, d \bar I_1 + \cdots + a_\tau^{IV} \, d \bar I_4. $$
Now, however, the coefficients $a$ in (11) are constants by assumption.

\newpage

\noindent 590 \hfill On the Erlangen program \hfill
\bigskip

\noindent We will therefore have the corresponding substitution formulas for our integrals $I_\tau$ themselves:
\begin{equation}%12
I_\tau = a_\tau^I \, \bar I_1 + \cdots + a_\tau^{IV} \, \bar I_4.
\end{equation}
(and of course also for the $I_\tau^*$), with which the result to be deduced here is already achieved.

The intellectual progress, however, which is connected with these formulas (12), can be stated: the $dI_\tau$, $dI_\tau^*$ are like all vectors in the general theory of transformation, by definition each is bond to a certain world point $w$ as a starting point, they are {\it bound} vectors (or, if we want to express ourselves more precisely: four-vectors). This attachment to a particular point now occurs in the transformation formulas for the $I_\tau$, $I_\tau^*$ all the way back.
\textit{One will suitably regard the $I_\tau$, $I_\tau^*$ as referring to free, compliant four-vectors},
i.e. as four-vectors, having only one direction and one intensity
$(= \sqrt{\sum g^{\mu\nu} I_\mu I_\nu})$, but have no particular place in the four-dimensional world.

Of course, this concept of the free four-vector is very much what we based as the group (11) of the affine transformations of the $w$. In physics, especially mechanics is just like I stated in my Erlangen program for geometry: that is a distinction of certain size types can only be talked about if you have agreement of the transformation group, which we take to clarify the concept formation. I have been advocating for decades that the physicists here want the underlying concept, which alone creates clarity.\footnote{
Comp. among others my essay ``The Screw Theory of Sir Robert Ball'' in 47th volume of the journal for math, and physics (1902), (1906 in the 62nd volume of Math. Annals reprinted with some extensions). [S. Dep. XXIX of this edition.] (As in the text, new physical concepts are not introduced there, but it will only be that which deals with the individual problems made by many, based on a clear mathematical principle.)}
In particular, I have in 1910 in my presentation on the geometric foundations of the Lorentz group\footnote{
Annual Report of the German Mathematical Society, vol. 19 (1910), prints in the Physikalische Zeitschrift, 12th year, 1911. [p. Dep. XXX of this statement.]}
expressly stated that you should never talk about relativity theory absolutely, but always only of an invariant-theory relative to a group. --- There are as
many kinds of relativity theory as there are groups.\footnote{
Compare also the message about ``Invariant variational problems'' of Miss Noether in the year 1918, G\"ottinger Nachrichten (final remark there).}

\newpage

\hfill XXXIII. Integral form of conservation laws, etc. \hfill 591
\bigskip

The view expressed in this way may be in contrast to the
disputes, as they have often been propagated in the wake of Einstein's general expositions,
not that I would like to emphasize or attach importance to Einstein's own individual developments.
Rather, the Einstein work, which I comment on in the present note that Einstein in the individual case --- without
systematically conceiving the concepts --- has served exactly the freedom of forming ideas, as I have recommended in my Erlanger program.

\bigskip
\begin{center}
\S 2.

\medskip
\textbf{The integrals $I_\tau$, $I_\tau^*$ for closed systems.}
\end{center}
\bigskip

Einstein understands a ``closed'' system in his above message mentioned under 3), which, so to speak is ``floating'' in a Minkowski world, i.e. a system whose individual parts go through a world-tube \emph{outside of which} there is a $ds^2$ of vanishing Riemannian curvature. You can do this by writing $ds^2$ with constant coefficients (without it being in the typical form of having to set $dt^2 - \frac{dx^2 + dy^2 + dz^2}{c^2}$): Einstein then speaks of ``Galilean'' coordinates.  As such, the $w^\varrho$ henceforth are chosen outside the world-tube, inside they may run arbitrarily, provided the transition is continuous. About the values $\frak{V}_\tau{}^\sigma$, $\frak{V}_\tau{}^{*\sigma}$
accordingly, nothing special can be said inside the tube but outside they are zero anyway. Because not only all $\frak{T}_\tau^\sigma$ vanish there but also, because of the constancy of $g_{\mu\nu}$ --- as a look at the definition formulas (5), (6) shows---, so do also all the $\frak{U}_\tau{}^\sigma$ or $\frak{U}_\tau{}^{*\sigma}$.

We think of the inside of the world-tube, the points of the system accordingly, furrowed of course, from a continual multitude of world lines all of which have to have a common positive sense.
Any vector tangent to the world line that marks that sense, may have the components $dw^I, \cdots, dw^{IV}$.

It is obvious which three-dimensional manifolds (hypersurfaces) will be referred to as ``cross sections'' Q of the world-tube.
In order to be able to express ourselves more comfortably, we will subsequently, consider only such cross sections, that are cut in {\it only} one point by each world line. Three mutually independent vectors $d', d'', d'''$, which are tangent to the cross-section, can then be chosen like this so that the determinant
$$ \left| \begin{array}{ccc}
            dw^I & \dots & dw^{IV} \\
            d'w^I & \dots & . \\
            d''w^I & \dots & . \\
            d'''w^I & \dots & .
          \end{array} \right| $$

\newpage

\noindent 592 \hfill On the Erlangen program \hfill
\bigskip

\noindent receives a fixed sign. Using the example:
$$ \begin{array}{cccccc}
    d & = & 0, & 0, & 0, & dt \\
    d' & = & dx, & 0, & 0, & 0 \\
    d'' & = & 0, & dy, & 0, & 0 \\
    d''' & = & 0, & 0, & dz, & 0
  \end{array} $$
if we follow the example, we will negatively choose this sign.

Assuming this, we form for the cross section the four integral
$$ I_\tau, \quad \hbox{or} \quad I_\tau^* $$
of the previous paragraph.

These developments can be see in close connection to Einstein. We then make the assertion, \emph{that these integrals are independent from the selection of cross sections, as well as from the coordinate choice that we might select inside the tube}. From the point of view which the whole world affine transformations of $w$ the $I_\tau$ or $I_\tau^*$ define in any case, a free, contragedient vector. The new expressions of Einstein state \emph{that these vectors are only of the material system as such, but are not dependent on the coincidences of analytical representation}.

In any case, to prove the new propositions, it suffices to place two such
cross-sections next to each other, $Q$ and $\bar Q$ taken together
delineate a uniform piece of the world-wide tube (i.e. do not intersect each other); --- the general case where $Q$ and $\bar Q$ interpenetrate,
is settled afterwards with ease by having a third
cross-section $(Q)$, which meets neither $Q$ nor $\bar Q$, and now
first compose $Q$ with $(Q)$, then $(Q)$ with $\bar Q$.

For the rest, the proof is divided (all following Einstein)
in two parts:

a) We first think of the coordinate system of $w$ within
and outside of the world tube somehow chosen by a rule. We
then think about the tube section between $Q$ and $\bar Q$ continuously rounded
on the outside, giving it a uniform hypersurface
which appears bounded, which penetrates the inside of the tube in $Q$ and $\bar Q$.
The integrals $I_\tau$, $I_\tau^*$ are, mutatis mutandis, over this closed hypersurface
extended, according to the previous paragraph, all zero.
But those provide parts of our hypersurface that protrude beyond the world tube to these integrals --- because for them the integrands $\frak{V}_\tau^\sigma$, $\frak{V}_\tau^{*\sigma}$, themselves vanish --- making no contribution whatsoever.

Remaining are the contributions of two cross sections $Q$ and $\bar Q$, which, however, if we calculate them according to the previously agreed sign rule, enter in the over the closed hypersurface integral with the opposite sign.
Since the sum is zero, the mentioned contributions each other are the same, QED.

\newpage

\hfill XXXIII. Integral form of conservation laws, etc. \hfill 593
\bigskip

b) Now it is important to understand that the $I_\tau$, $I_\tau^*$ relevant to the individual cross sections $Q$ in all modifications of the $w^\varrho$ which vanish outside the world's tube, in fact remain unchanged.
We do this in a way that we first inside the tube think of two coordinate systems, $w$ and $\bar w$, given on the
band of the tube both connect in a smooth manner to the same outer (Galilean) coordinate system. We use the former to calculate the integrals $I_\tau$, $I_\tau^*$ for the cross-section $Q$ and the latter to calculate for $\bar Q$, where the values $\bar I_\tau$, or $\bar I_\tau^*$ result.
It's being closed
shows that $I_\tau = \bar I_\tau$, respectively $I_\tau^* = \bar I_\tau^*$
and this proof is furnished
if we succeed in introducing a third coordinate determination, $\bar{\bar w}$, which extends along $Q$ with sufficient accuracy to that of $w$, along $\bar Q$
likewise to which the $\bar w$ adjoins, while along the mantle of the
tube and outside the same there still prevails
Galileo coordinates. ``Sufficiently accurate'' means that
the calculation of $V_\tau^\sigma$ or $V_\tau^{*\sigma}$ from the $\bar{\bar w}$ for the cross-section $Q$
gives the same results as the use of $w$ and accordingly
for the cross section $\bar Q$ the same results as the use of the $\bar w$.
Because of the differential quotients of $g_{\mu\nu}$ in the formulas (5), (6) in the definition of $V_\tau^\sigma$, $V_\tau^{*\sigma}$, it is sufficient in this respect,
--- after a result Mr Vermeil obtained for me --- that
the $\bar{\bar w}$ with the $w$ along $Q$ also agree in their three first differential quotients, likewise with the $\bar w$ along $\bar Q$. All the conditions imposed on the determination of the coordinates $w$ are now apparently satisfied by the following example: Let us introduce the {\it equations} for which the cross sections $Q$, $\bar Q$ or in the $\bar w$ and the $w$ suffice. Let $f(\bar w) = 0$ be the first of these equations, $\bar f(w) = 0$ the second.
I'll just write:
\begin{equation}%13
\bar{\bar w} = \frac{\left( \bar f(w) \right)^4 \cdot w + \left( f(\bar w) \right)^4 \cdot \bar w}{\left( \bar f(w) \right)^4 + \left( f(\bar w) \right)^4}
\end{equation}
and indeed they have complied with all the conditions, so our second proof is furnished, and thus the proof of the new statement is done absolutely.

\bigskip
\begin{center}
\S 3.

\medskip
\textbf{Final determination of free momentum energy vectors for the closed system.}
\end{center}
\bigskip

The $I_\tau$ and $I_\tau^*$ form of course the basis for the momentum energy vectors to be added to the closed system. To fully define the latter, however, it will still be necessary to consider the dimensions of the interconnected types of quantities.

\vfill
{\tiny{Klein, Gesammelte math. Abhandlungen. I.}}\hfil 38 \hfil

\newpage

\noindent 594 \hfill On the Erlangen program \hfill
\bigskip

\noindent On page 569 of my previous note, it was agreed that the dimension sec$^2$ should be added to $ds^2$. Accordingly, let us assume that the $w^\varrho$ used all have the dimensions sec$^{+1}$. The $g^{\mu\nu}$, $g_{\mu\nu}$, and the $g$ are then dimensionless, and the $K$, $U_\tau^\sigma$, $\frak{U}_\tau^\sigma$, $U_\tau^{*\sigma}$ $\frak{U}_\tau^{*\sigma}$ matching the dimension sec$^{-2}$. Since the gravitational constant $\chi$ has the dimension gr$^{-1}$ cm$^{+1}$, for the $\frac{{\textstyle\mathfrak{U}}{}_\tau^\sigma}{\chi}$ $\frac{{\textstyle\mathfrak{U}}{}_\tau^{*\sigma}}{\chi}$ we obtain the dimension gr$^{+1}$ cm$^{-1}$ sec$^{-2}$, that is, the dimension of a ``specific'' (to the spatial unit related) energy. This is valid as they join together in the $\frak{V}_\tau^\sigma$, $\frak{V}_\tau^{*\sigma}$ additively with the $\frak{T}_\tau^\sigma$.

Now these $\frak{V}_\tau^\sigma$, $\frak{V}_\tau^{*\sigma}$ under the integral signs $I_\tau$, $I_\tau^*$ are multiplied by tripartite determinants which, according to our agreement on the dimension of $w$, themselves have the dimension sec$^{+3}$. Apparently, in order to get the dimension of an actual energy, I have to $I_\tau$, $I_\tau^*$ still add the factor $c^3$ [$c$ = speed of light]. \emph{In accordance herewith, the magnitude quadruples shall be definitively designated as free momentum energy vectors of the presented closed system}:
\begin{equation}%14
J_\tau = c^3 I_\tau, \qquad J^*_\tau = c^3 I_\tau^*.
\end{equation}
Numerical factors that may still be doubtful should not be further attached; also one should be held to our sign determination.

I see the proof of the correctness of this approach in the fact that in the definition of our $J^*_\tau$ Einstein's definition of the momentum energy vector belonging to the closed system is included. In order to understand this we shall first of all, in our definition of $J_\tau^*$ strike off the factor $c^3$ (because Einstein uses such units of measurement assuming that $c = 1$, which is a mere externality for the comparison in question). But then we have to do what is a real particularization, choose the cross-section $Q$ so that the freedom of choice of coordinates left to us by the equation $w^{IV} = 0$ can be represented. Considering the extent of this limitation, consider that the Galilean coordinates outside the world tube are set to be an affine transformation. The new condition is therefore to choose the cross-section $Q$ so that it intersperses the mantle of the world-tube of our system in a structure which, seen from the outside with initially arbitrary assumed Galileo-altitude coordinates, is represented by a \emph{linear} equation.

\newpage

\hfill XXXIII. Integral form of conservation laws, etc. \hfill 595
\bigskip

Let's assume, in fact, that along the cross section
$w^{IV}$ disappears, that is to say, $d'w^{IV}$, $d''w^{IV}$, $d'''w^{IV}$ are in fact zero. Our
integral $I_\tau^*$  then reduces (by setting $c = 1$) to
$$
\int\!\int\!\int \frak{V}_\tau^{*4} \left|
\begin{array}{ccc}
  d'w^I & \cdots & d'w^{III} \\
  d''w^I & \cdots & d''w^{III} \\
  d'''w^I & \cdots & d'''w^{III}
\end{array} \right|
$$
so, if we go back to the ordinary transcription,
\begin{equation}%15
J_\tau^* = \int\!\int\!\int \frak{V}_\tau^{*4} dw^I dw^{II} dw^{III},
\end{equation}
which is exactly the Einstein formula except for the letters.

This formula is indeed, outwardly, undoubtedly simpler than the one I used. For this is then the vector character of
the $J_\tau^*$, as Einstein claims but does not substantiate in more detail, being more difficult to see. In lengthy correspondence with Einstein
I originally did not want to succeed in establishing this vector character until I saw the Grassmann transcription
for the integral expression that I used above.
But that was also the generalization of the cross-sectional concept I chose to give.

There remains the essential difference regarding the Einstein representation, that in addition to the vector $J_\tau^*$ with equal right I place the vector $J_\tau$, ---  under the integral sign, instead of Einstein's
$\frak{t}_\tau^\sigma = \frac1{\chi}\frak{U}_\tau^{*\sigma}$ set the Lorentz $\frac1{\chi} \frak{U}_\tau^\sigma$.
\emph{That the $J_\tau$ and the $J_\tau^*$ are in general different, we will immediately see from an example}. After that I could even get an infinite number, assigning to the completed system
many different momentum energy vectors, if I e.g. instead of $\frak{t}_\tau^\sigma$ put the aggregate $\frak{t}_\tau^\sigma + \lambda \left( \frac1{\chi} \frak{U}_\tau^\sigma - \frak{t}_\tau^\sigma \right)$, with $\lambda$ any numerical constant.
Anyway, instead of the $\frak{t}_\tau^\sigma$ one can be allowed to place any $\frak{U}_\tau^\sigma$ that differs from the $\frak{t}_\tau^\sigma$ only by a term of the dimension required, which represents a mixed tensor compared to affine transformations that vanishes identically outside the world-tube, but inside has vanishing divergence. Which of these infinite multitudes of vectors is preferable as long as I only demand the integral statements remains undecided.
A decision can only be made if you bring up new reasons that determine of the infinite many forms of the differential just a single preferred one.

\vfill

\hfill $38^*$ \qquad\qquad

\newpage

\noindent 596 \hfill On the Erlangen program \hfill
\bigskip

\begin{center}
\textbf{II. Einstein's spatially closed world (cylinder world).}

\medskip
\S 4

\medskip
\textbf{The closed space of constant positive curvature.}
\end{center}
\bigskip

In Einstein's note of February 1917 there is only the possibility
of a {\it spherical} space, as he conceived of a manifold of
four dimensions $(\xi,~\eta,~\zeta,~\omega)$ whose arc element is given by the equation
\begin{equation}%16
d\sigma^2 = d\xi^2 + d\eta^2 + d\zeta^2 + d\omega^2,
\end{equation}
being cut out directly by the ``sphere equation''\footnote{From now on, ``space'' is understood to mean a three-dimensional area
(which is contained in the four-dimensional ``world'').}
\begin{equation}%17
\xi^2 + \eta^2 + \zeta^2 + \omega^2 = R^2.
\end{equation}
For the connoisseur of geometric literature it is, of course, that I immediately put Einstein onto my old
attentions of 1871 made on non-Euclidean geometry of 1871, according to which, in addition to the spherical spatial form another
closed spatial form of constant positive curvature, the elliptical space (as it of me in connection with my other considerations
was called at that time). You get it from the spherical space by simply identifying two diametrically opposed points of the sphere
by a central projection onto a contacting linear space. We like to put accordingly:
\begin{equation}%18
x = R \frac{\xi}{\omega}, \quad y = R \frac{\eta}{\omega}, \quad z = R \frac{\zeta}{\omega},
\end{equation}
inversely it then becomes:
\begin{equation}%19
\xi = \frac{R x}{\sqrt{x^2 \!+\! y^2 \!+\! z^2 \!+\! R^2}}, \quad \eta = \cdots, \quad \zeta = \cdots, \quad \omega = \frac{R^2}{\sqrt{x^2 \!+\! y^2 \!+\! z^2 \!+\! R^2}}.
\end{equation}

The elliptical space is simpler than the spherical one by itself, its geodesic lines are simply {\it straight} lines (if they meet at all, they only cut in {\it one} point\footnote{
That is why the elliptical space precedes, if one, as I did in 1871, starting from the basic concepts of projective geometry. It is then with the hyperbolic spaces (the spaces of Bolyai and Lobachevsky), like the parabolic spaces (the Euclidean spaces) directly adjacent, and it means to thoroughly misjudge this relationship if, as the majority of authors say, formulas (18) are used as a ``projection'' of the spherical space on the ``Euclidean''. ``Euclidean'' is the sum of the value systems of three variables $x, y, z$ only if we add the differential form $dx^2+ dy^2 +dz^2$, or --- for the group-theoretical order if we consider the totality of the projective transformations of the $x, y, z$ (whose invariant theory is projective geometry) through the subgroup of those replace, which leave the said differential form unchanged. --- I bring all these things, otherwise well-known, in the present work, which is also intended for physicists, as a matter of language, because they still do not seem to be widespread among physicists under the influence of the one-sided Helmholtz tradition dating back to 1868.});
the length

\newpage

\hfill XXXIII. Integral form of conservation laws, etc. \hfill 597
\bigskip

\noindent  of such a geodesic line is $R \pi$, the total content of the space is $R^3 \pi^2$ (instead of $2 \pi R$, or $2 R^3 \pi^2$ in the spherical cases).

In the mere statement of the arc element, the difference occurring of course is not yet apparent in both spatial forms.\footnote{
With the indication of $d\sigma^2$ is in fact the ``context'' that the associated spatial form in the large shows, being not yet determined. This too is still too often ignored in the contemporary literature. For spaces of constant curvature I have treated in detail the relevant conditions in a treatise of 1890 [Math. Annalen, vol. 37 (see treatise XXI of this edition)]. Of textbooks see in particular that of Killing (Introduction to the Foundations of Geometry, Part I, 1893). I also refer gladly to the recent publications by Hadamard and Weyl.}
I can use the $d\sigma^2$ given through (16) and (17) just as well for the elliptical space as its value recalculated into $x, y, z$:
\begin{eqnarray}%20
d\sigma^2 &\!=\!& \frac{R^2}{(x^2 + y^2 + z^2 + R^2)^2} \Bigl\{ R^2 (dx^2 + dy^2 + dz^2) + (y dz - z dy)^2 \\
&\!\!& \qquad + (z dx - x dz)^2 + (x dy - y dx)^2 \Bigr\} \nonumber
\end{eqnarray}
in the spherical case, or in both cases the value can be expressed in polar coordinates:
\begin{equation}%21
d\sigma^2 = R^2 \left( d\vartheta^2 + \sin^2\vartheta \cdot d\varphi^2 + \sin^2\vartheta \sin^2\varphi \cdot d\psi^2 \right).
\end{equation}

\bigskip
\begin{center}
\S 5.

\medskip
\textbf{Einstein's ``cylinder world'' and its group.}
\end{center}
\bigskip

Furthermore, as in the previous paragraph, $d\sigma^2$ --- also without specifying the coordinate system --- in short, be the square of the arc element of a closed space of the constant curvature $\frac1{R^2}$  may now be assumed to be spherical or elliptical. The rise to Einstein's spatially closed world will then just do that, that we set
\begin{equation}%22
ds^2 = dt^2 - \frac{d\sigma^2}{c^2}
\end{equation}
and by the way let $t$ from $-\infty$ to $+\infty$ (excluding these limits) run. (The dimension and sign of this $ds^2$ agree with our general conventions. If we then formally calculate the curvature measure for the space $t = $const., we obtain $-\frac{c^2}{R^2}$. Of course, this

\newpage

\noindent 598 \hfill On the Erlangen program \hfill
\bigskip

\noindent negative sign corresponds only to the circumstance that the $ds$ introduced in (22) becomes purely imaginary for the space mentioned; so it is no contradiction to the previous paragraph, where we have briefly referred to space as such having a constant positive curvature.)

We ask first and foremost for the largest continuous group of coordinate transformations, through which the $ds^2$ (22) into itself passes.

It is clear from the beginning that at least one such transformation $G_7$ exists. For there is already a continuous $G_6$, which takes
$d\sigma^2$ into itself: to connect to (16), the epitome of the orthogonal transformations of $\xi, \eta, \zeta, \omega$ of  determinant +1. You
then include $G_1$ which corresponds to an increase of $t$ by an arbitrary constant.  The $G_7$ thus obtained is certainly transitive,
i.e. through them you can transfer each world point into every other, for example into the point $t = 0$, $\vartheta = 0$ (to make use of the polar coordinate system introduced in (21)). This
point may be briefly called $O$; around it there is still a continuous $G_3$
possible from space rotations.

We now claim, \emph{that there is no larger continuous group of coordinate transformations that converts $ds^2$ into itself, as our $G_7$}. For the purpose of this it suffices to show that when $O$ is held fixed just the mentioned $G_3$ consists of rotations. ``Riemann normal coordinates'' emanating from $O$ provide evidence.
This is achieved, for example, by maintaining $t$ as a variable and instead
of the polar coordinates $\vartheta, \varphi, \psi$ introduce the connections:
\begin{equation}%23
y_1 = \frac{R}{c} \vartheta \cdot \cos\varphi, \quad y_2 = \frac{R}{c} \vartheta \cdot \sin\varphi \cos\psi, \quad y_3 = \frac{R}{c} \vartheta \cdot \sin\varphi \sin\psi.
\end{equation}
Let us still write for $t$ for uniformity $y_4$, thus we obtain for $ds^2$
\begin{eqnarray}%24
ds^2 &=& (dy_4^2 - dy_1^2 - dy_2^2 - dy_3^2) + \frac{c^2}{3 R^2} \sum_{1,2,3} (y_i dy_\chi - y_\chi dy_i)^2 \\
&& + \hbox{ higher order terms in the } y_1,~y_2,~y_3, \nonumber
\end{eqnarray}
which shows that we are indeed dealing with normal coordinates.
As for the transformations of $ds^2$ into itself, we have since $O$ should remain fixed, according to the general theory of normal coordinates only after the largest continuous group to ask for \emph{homogeneous linear} substitutions of $y$, which transforms this $ds^2$ into itself. The two written terms of the $ds^2$ must, for their dimensions, each pass over to themselves. It is clear then that $y_4$ must remain unchanged, while $y_1$, $y_2$, $y_3$ can

\newpage

\hfill XXXIII. Integral form of conservation laws, etc. \hfill 599
\bigskip

\noindent be subjected at most to the continuous group of ternary orthogonal substitutions with determinant 1. But with that we are already at the goal.

According to the theorem thus proved, it may be allowed to refer to Einstein's spatially closed world as the \emph{cylindrical world}, because
it has, so to speak, the symmetry of a rotary cylinder: any displacement along the $t$-axis and any rotation around $O$ while holding $t$ fixed. Of course, the analogy is not perfect because it can be rotated about any point other than $O$. I also do not want to introduce a permanent term, but only to ad hoc have a short expression, which denotes the opposite of de Sitter's hypothesis labeled $B$ to be treated in the next section.

For the rest, we shall say that in this case, after we have agreed on the time unit and the starting point of the time, \emph{the concept of time does not contain any arbitrariness}\footnote{
So also has de Sitter loc.\ cit.\ noted.},
or, if one prefers to say so, that within the four-dimensional world \emph{the threefold extended spaces $t = $const.\ manifolds are unique}. So this is a remarkable approach to the ways of imagining classical mechanics.

This is, considering the physical consideration of the cylinder world introduced by Einstein, naturally from the outset. In order to overlook the totality of the mass distributions and events of the world from the higher point of view, Einstein first imagined a state of dissection in which the totality of the masses in the space, which is defined to be closed, is \emph{incoherent}, and \emph{uniformly distributed}, and \emph{at rest} within this space, while $t$ runs from $-\infty$ to $+\infty$. The actual mass distributions and incidents should be interpreted as deviations from this average state.
Measured at this average state is then the time (or more precisely the time difference of two world points measured in an agreed unit) in fact something absolute, the space is itself homogeneous\footnote{
The fact that the space can be assumed at will to be spherical or elliptical Einstein had at that time approved without further ado. Incidentally, de Sitter also treats these two assumptions side by side. Likewise also the new Weyl book (space, time, matter)}.
However, this conception finds its precise mathematical expression in the invariant theory of our $G_7$.

It is particularly interesting to see how our $G_7$ to the \emph{Lorentz group} $G_{10}$ is expanded, so one can see how the \emph{ideas of the ``special'' theory of relativity} come when the measure of the curvature of our space vanishes, i.e. setting $R=\infty$. Our

\newpage

\noindent 600 \hfill On the Erlangen program \hfill
\bigskip

\noindent $ds^2$ (22) is reduced then in fact to its first term\footnote{
Not only the second term, but also all higher terms are dropped.}:
$dy_4^2 - dy_1^2 - dy_2^2 - dy_3^2$ and then remains with all homogeneous linear substitutions of the $dy_1$, $dy_2$, $dy_3$, $dy_4$ unchanged, to which this {\it single} square shape transforms into. \emph{Thus $y_4 = t$ ceases to be a stand-alone variable, rather, it combines with the permissible substitutions with the $y_1$, $y_2$, $y_3$, as this is just the very nature of the special relativity theory}.

\bigskip
\begin{center}
\S 6.

\medskip
\textbf{The field equations of the cylinder world}
\end{center}
\bigskip

We still have to confirm that the assumption the whole space uniformly filled with resting matter, say of the
constant density $\varrho$, is indeed compatible with our $ds^2$ which we put up for Einstein's field equations. Of course the thought is of the field equations ``with $\lambda$-term'', of which already in my previous note (formula 57) were mentioned:
\begin{equation}%25
K_{\mu\nu} - \lambda g_{\mu\nu} - \chi T_{\mu\nu} = 0.
\end{equation}
Because the distribution of matter in space is supposed to be quite uniform it is sufficient to make the verification for the point $O$. Also
we will, since it is a relation acting between tensor components, from the outset the $ds^2$ (24) written in normal coordinates may be taken as a basis. But from here you can find without any special calculation, cf. the note from Vermeil in the G{\"o}ttinger Nachrichten of October 26, 1917
(``Note on the mean curvature of an-fold expanded Riemannian manifold''):
\begin{equation}%26
K_{11} = K_{22} = K_{33} = -\frac{c^2}{R^2}, \quad K_{44} = \frac{3c^2}{R^2},
\end{equation}
while all the other $K_{\mu\nu}$ vanish.

Now you have for the point $O$ on the basis of the normal
coordinates:
\begin{equation}%27
\hbox{all} \quad T_{\mu\nu} = 0, \quad \hbox{except for} \quad T_{44} = c^2 \varrho.
\end{equation}
Therefore the field equations (25) result in:
$$ - \frac{c^2}{R^2} + \lambda = 0, \qquad \frac{3 c^2}{R^2 } -\lambda - \chi c^2 \varrho = 0, $$
i.e.,
\begin{equation}%28
\lambda = \frac{c^2}{R^2}, \qquad \varrho = \frac{2}{\chi R^2},
\end{equation}

\newpage

\hfill XXXIII. Integral form of conservation laws, etc. \hfill 601
\bigskip

\noindent which agree with the results given by Einstein himself (as long as
one still sets $c^2 = 1$).

Here the remark is that for $K$ itself the following constant
value is calculated:
\begin{equation}%29
K = \frac{6 c^2}{R^2}.
\end{equation}

For application to the universe, of course, it remains from our present knowledge of stellar astronomy with some probability
to estimate the corresponding value of $R$. This de Sitter has executed in his repeatedly mentioned communication. I like to present his result to see that Einstein's cosmological view, whose mathematical content alone we treat here, but also physically does not completely hang in the air. You have to go to de Sitter to take
$$ R = 10^{12} \ \hbox{ to } 10^{13} \hbox{ half of Earth's orbit}. $$
The density $\varrho$ is so low that there is only about $10^{-26}$ gr to the cubic centimeter, i.e., in about 100 cubic centimeters has the mass of a hydrogen molecule. The constant $\lambda$ however becomes casually $10^{-30}$ sec$^{-2}$.

\bigskip
\begin{center}
\S 7.

\medskip
\textbf{The integral statements for the cylinder world.}
\end{center}
\bigskip

If one takes the field equations with $\lambda$-element, then, as I have declared in \S 7 of my previous note following Einstein's developments, $U_\tau^\sigma$ and $t_\tau^\sigma = \frac1{\chi} U^*{}_\tau^\sigma$, so that the conservation laws are preserved by replacing
\begin{equation}%30
\bar U_\tau^\sigma = U_\tau^\sigma + \lambda \delta_\tau^\sigma, \quad
\bar t_\tau^\sigma = t_\tau^\sigma + \frac{\lambda}{\chi} \delta_\tau^\sigma.
\end{equation}
We accordingly instead of the integrals $I_\tau$, respectively $I^*_\tau$ of \S 1 integral $\bar I_\tau$ respectively $\bar I_\tau^*$ form and are sure from the outset that these integrals, taken over such closed hypersurfaces, which delimit part of the cylinder world, vanish.

Now, the concept of the \emph{cross-section}, which we used in $I$ for the ``world-tube'' considered at that time will have to be transferred. We will want to denote any other closed hyperface as such, for which every world line of the cylinder world, i.e. every parallel to the $t$-axis
cuts once. The simplest example is the ``spaces'' $t = $ constant.

We will then have the double statement as before:

1. that the integrals $I_\tau$ or $I_\tau^*$, taken for any cross section, have a value independent of its selection;

\newpage

\noindent 602 \hfill On the Erlangen program \hfill
\bigskip

2. that this value does not depend on which coordinates one uses in the execution of the cross-section extended integration.

The only thing that will change, is that it is no longer allowed, that the epitome of the integral $\bar I_\tau$, or $\bar I_\tau^*$ designates a (free) four-vector. Because of the nature of our $G_7$, the group-theoretical basis for this labeling is lacking. In any case:

$I_4$ or $\bar I_4^*$ will naturally stand out for themselves.  We may call the value, multiplied by $c^3$, the \emph{total energy of the cylinder world}.

But by the classification of the quantities $\bar I_1$, $\bar I_2$, $\bar I_3$, (or the $\bar I_1^*$, $\bar I_2^*$, $\bar I_3^*$) we do not have much to worry about, since one can convince oneself in various ways \emph{that they are all zero}.

Firstly, this follows (as Einstein also states for $\bar I^*_\tau$) on symmetry grounds. I summarize the matter from my point of view. Of course, if we hold to the normal coordinates $y$, then of the $\infty^6$ continuous transformations which convert the space $y_4 = 0$ into themselves, only the $\infty^3$ are represented as homogeneous linear substitutions of the $y_1$, $y_2$, $y_3$ which rotate the space imagined around $O$.  But this is sufficient for our purposes, too, the ones they have formed to consider a subgroup. With respect to it the $\bar U_1^\sigma$, $\bar U_2^\sigma$, $\bar U_3^\sigma$ will behave (as well as the $\bar U_1^{*\sigma}$, $\bar U_2^{*\sigma}$, $\bar U_3^{*\sigma}$) like the components of a three-dimensional tensor, that is, the $\bar I_1$, $\bar I_2$, $\bar I_3$ (and the $\bar I^*_1$, $\bar I^*_2$, $\bar I^*_3$), behave as the components of a threefold vector running out of $O$.
Now however the cylinder world is, as we know, spatially isotropic around $O$. Said three-vector must therefore remain unchanged for any space rotation around $O$, and it can only do that if all of its components vanish.

Second, we may take the direct calculation route. We choose as the cross section over which our integrals are to extend any of the manifolds $y_4 = $ const. Within those, any coordinates $w^I$, $w^{II}$, $w^{III}$  are thought to be introduced. The integrals $\bar I_\tau$ and $\bar I_\tau^*$  then, according to the statements of \S 3, in the abbreviated form can be written:
\begin{equation}%31
{\bar I}_\tau = \int\!\int\!\int \Bigl( T^4_\tau + \frac1{\chi} \bar U^4_\tau \Bigr) \sqrt{-g} \cdot dw^I dw^{II} dw^{III}
\end{equation}
or \\
%31*
$$ (31^*) \hspace{2cm} {\bar I^*}_\tau = \int\!\int\!\int \Bigl( T^4_\tau + \bar t^4_\tau \Bigr) \sqrt{-g} \cdot dw^I dw^{II} dw^{III}. \hspace{3cm} $$
A direct calculation shows that $T^4_\tau$, $\bar U^4_\tau$, $\bar t^4_\tau$ for $\tau = 1, 2, 3$ all vanish.

\newpage

\hfill XXXIII. Integral form of conservation laws, etc. \hfill 603
\bigskip

For the \emph{total energy of the cylinder world} we have obtained the expressions with these formulas:
\begin{equation}%32
\bar J_4 = c^3 \int\!\int\!\int \Bigl( T^4_4 + \frac1{\chi} \bar U^4_4 \Bigr) \sqrt{-g} dw^I dw^{II} dw^{III}
\end{equation}
or
$$ (32^*) \hspace{2cm} \bar J^*_4 = c^3 \int\!\int\!\int \Bigl( T^4_4 + t^4_4 \Bigr) \sqrt{-g} dw^I dw^{II} dw^{III}. \hspace{3cm} $$
The amount of energy then appears in the one and the other case as the sum of two summands. --- We may refer to the summand corresponding to $T^4_4$ as the \emph{mass energy}, to the other as the \emph{gravitational energy}.

The mass energy is now calculated easily. Namely $T^4_4$, whatever we may choose $w^I$, $w^{II}$, $w^{III}$, equals $c^2 \varrho$, and $c^3 \sqrt{-g} dw^I dw^{II} dw^{III}$ is nothing else than the volume element $d V$ of our space $y_4 =$ const. \emph{Thus, the mass energy is simply $c^2\varrho V$}, where $V$ is the total volume of space, that is, depending on the spherical or elliptic hypothesis, $2 \pi^2 R^3$  or $\pi^2 R^3$.

\emph{But for the gravitational energy Einstein has in his case}, that is, on the basis of the formula ($32^*$), using spatial polar coordinates, \emph{found zero}. In this case the $dV$ becomes $= \sin^2\vartheta \sin\varphi \cdot d\vartheta d\varphi d\psi$, the ${\bar t}^4_4$ (if I contract the Einstein terms) $\frac{\cos2\vartheta}{\sin^2\vartheta}$, the result of integration becomes zero, because $\int \cos2\vartheta\cdot d\vartheta$ is to be taken from 0 to $\pi$.  --- This result is certainly very remarkable. Since it must be independent of the choice of $w^I$, $w^{II}$, $w^{III}$, one wonders if, instead of the polar coordinates, which entail a longer mechanical calculation (only hinted by Einstein), one should not usefully introduce others. I would like to suggest to operate consistently with the supernumary coordinates $\xi$, $\eta$, $\zeta$, $\omega$  of \S 4 (between which then the dependence $\xi^2 + \eta^2 + \zeta^2 + \omega^2 = R^2$ exists). Of course, one has to generalize the basic formulas of tensor calculations to the case of dependent coordinates, for which, however, all the approaches in the literature are available. I suspect that when performing this implementation not only the integral of the gravitational energy of all the volume elements of the space, but already the differential, corresponding to the individual volume element would vanish, from which nevertheless an improved insight into the simplicity of Einstein's result would be achieved.

\newpage

\noindent 604 \hfill On the Erlangen program \hfill
\bigskip

So much for the $\bar t^4_4$. The new thing I have to do now is \emph{that we get a completely different result} (and this without any complicated calculation), \emph{if we select instead of the $\bar t^4_4$ the $\frac1{\chi} \bar U^4_4$, that is, in place of $\bar J_4^*$ choose the $\bar J_4$}.  --- $\bar U^4_4$ is, as we know $= U^4_4 +\lambda$. If we now, come back to the formula for $U^4_4$ mentioned above under (5), it turns out that in the case of the cylinder world, with any choice $w^I$, $w^{II}$, $w^{III}$, all terms except the first are dropped. $U^4_4$ becomes simply $= \frac12 K$ and so
\begin{equation}%33
\bar U^4_4 = \frac12 K + \lambda = \frac{4c^2}{R^2}.
\end{equation}
So it has a constant but not vanishing value.  As a result, on the basis of $U^\sigma_\tau$ \emph{the gravitational energy of the cylinder world becomes not zero, but twice as large as the mass energy}.

The state of affairs hereby established clearly has achieved meaning that exceeds the case of the cylinder world. It shows by example that the energy components $\frac1{\chi} U^\sigma_\tau$ also for the \emph{integral forms} of the conservation laws give generally different results than the $t^\sigma_\tau$. This is what I mentioned in the introduction, the introduction of a subjective moment establishing the energy balance, and explained its scope for closed systems at the end of \S 3.  The result is certainly in no way marvelous, but it contradicts the impression that one has on the first reading of Einstein's note, as if it were an exclusive legal title for the $t^\sigma_\tau$ to lead to simple integral statements.

\bigskip
\begin{center}
\textbf{III. Regarding de Sitters hypothesis B.}
\end{center}
\bigskip

In his repeatedly mentioned communications, in particular Note 3 of the Monthly Notices, de Sitter has the assumption of the cylinder world, that he referred to as hypothesis A, among others modified so that he instead of the cylinder world --- while maintaining the characteristic key signature of $ds^2$  --- put a world of \emph{constant curvature}.  It is this hypothesis designated by him as B\footnote{
de Sitter notices that this assumption (which is familiar to the mathematician recommended by its symmetry) was first proposed by Ehrenfest. I have myself in my lectures from the spring of 1917 (whose elaboration has become public in a small number of copies), when I intended to refer to Einstein's then first published ``Cosmological Considerations'', but did not compare the formulas exactly, involuntarily made the same approach and then later, as I moved to work out the physics implications, of which, of course, the results did not agree with those given by Einstein for his cylinder world.};
I set myself the task to demonstrate convincingly the relationships occurring in this case by formulas that are as simple as possible.  Incidentally, the essence of my reflections can already be

\newpage

\hfill XXXIII. Integral form of conservation laws, etc. \hfill 605
\bigskip

\noindent
found in the minutes of the G\"ottingen meetings Mathematical Society of the summer of 1918 printed in the October 1918 Issues of the Annual Report of the German Mathematician union (oblique pages, pp. 42--44). See also a message to the Amsterdam Academy (Report from Sept. 29, 1918).

\bigskip
\begin{center}
\S 8.

\medskip
\textbf{The geometric foundations for the world of constant curvature.}
\end{center}
\bigskip

We will simply do justice to the assumption that the world is a manifold of constant curvature by writing the usual equation of a sphere in five variables with {\it a change of a sign}, and measure Euclidean on this ``pseudo-sphere''\footnote{
The precursory syllable ``pseudo'' should always indicate the occurrence of a different sign.}.
In the meantime, however, we want to keep the earlier assignment concerning the dimension of the variable, call the radius not $R$, but $\frac{R}{c}$; likewise, to be consequent, turn back the usual sign of $ds^2$. So I write as an equation of the pseudo sphere
\begin{equation}%34
\xi^2 + \eta^2 + \zeta^2 - \upsilon^2 + \omega^2 = \frac{R^2}{c^2}
\end{equation}
and for the associated $ds^2$:
\begin{equation}%35
-ds^2 = d\xi^2 + d\eta^2 + d\zeta^2 - d\upsilon^2 + d\omega^2.
\end{equation}

The resulting \emph{pseudospheric world} $(\xi, \eta, \zeta, \upsilon, \omega)$ has, because of the minus sign added to the $ds^2$, constant (Riemann) curvature $-\frac{c^2}{R^2}$. Moreover, it goes through  continuous $G_{10}$ ``pseudo-orthogonal'' substitutions, that is, by appropriate linear homogeneous substitutions of the $\xi,~\eta,~\zeta,~\upsilon,~\omega$ converts over into itself, but not, as one
can easily prove, by an even more comprehensive group. Then we put on our side a pseudo-elliptic world, writing the $ds^2$ given in (35):
\begin{equation}%36
x = \frac{R}{c} \cdot \frac{\xi}{\omega}, \qquad y = \frac{R}{c} \cdot \frac{\eta}{\omega}, \qquad z = \frac{R}{c} \cdot \frac{\zeta}{\omega}, \qquad u = \frac{R}{c} \cdot \frac{\upsilon}{\omega},
\end{equation}
from which inversely
\begin{eqnarray}%37
&& \xi = \frac{R x}{c \sqrt{x^2 + y^2 + z^2 - u^2 + \frac{R^2}{c^2}}}, \quad \eta = \quad , \quad \zeta = \quad , \quad \upsilon = \quad , \\
&& \qquad \omega = \frac{R^2}{c \sqrt{x^2 + y^2 + z^2 - u^2 + \frac{R^2}{c^2}}}. \nonumber
\end{eqnarray}

\newpage

\noindent 606 \hfill On the Erlangen program \hfill
\bigskip

\noindent We will be able to use these $\xi$, $\eta$, $\zeta$, $\upsilon$, $\omega$ in the treatment of the pseudo-elliptical world to homogenize the equations (as will often happen). Let us still note that
\begin{equation}%38
x^2 + y^2 + z^2 - u^2 + \frac{R^2}{c^2} = \frac{R^2}{c^2} \frac{\xi^2 + \eta^2 + \zeta^2 - \upsilon^2 + \omega^2}{\omega^2} = \frac{R^4}{c^4 \omega^2},
\end{equation}
is always positive, insofar as we restrict ourselves, of course, to real values of the original coordinates $\xi, \cdots, \omega$.

For the sake of brevity, we will now only talk of this pseudo-elliptical world (that is, leave the pseudospheric aside) and I have to ask the reader, to be allowed to use quite {\it projective} views which really do justice to the conditions in question. I want to in short compile a number in this regard of statements that are self-evident to the trained geometer:

1. In the pseudo-elliptical world, it is about a {\it projective dimensional determination}, whose fundamental structure is given by
\begin{equation}%39
x^2 + y^2 + z^2 - u^2 + \frac{R^2}{c^2} = 0
\end{equation}
and, according to the analogy, from now on may be briefly referred to as a (two-shelled) hyperboloid.
According to the sign definition (38), we find ourselves {\it between} the shells of this hyperboloid (that is, in the world-piece from which real tangential cones run to the hyperboloid), in accordance with the indefinite character of our $ds^2$.
In homogeneous coordinates $\xi$, \dots written is the equation of the hyperboloid:
\begin{equation}%40
\xi^2 + \eta^2 + \zeta^2 - \upsilon^2 + \omega^2 = 0,
\end{equation}
the hyperboloid is thus the intersection of the {\it asymptote cone} of our pseudo sphere with our $x$, $y$, $z$, $u$-region.

2. The continuous host of pseudo-orthogonal substitutions of the $\xi$, $\eta$ \dots returns the largest continuous group for the $x$, $y$, $z$, $u$ of collineations through which our hyperboloid merges into itself.

3. May we simply call \emph{spaces} the new entities, which are represented by a single linear equation between the $x$, $y$, $z$, $u$ (or by a corresponding homogeneous equation between the $\xi$, $\eta$, \dots).

4. Spaces containing the fundamental hyperboloid only in imaginary cut points (such as $u = 0$) will absolutely exhibit elliptical measures, and thus be of limited extension. In that respect one comes to designate our world as ``spatially closed'' and directly place it next to the Einstein cylinder world.

\newpage

\hfill XXXIII. Integral form of conservation laws, etc. \hfill 607
\bigskip

5. In addition to these spaces, borderline cases are those that touch the hyperboloid in a point, e.g. the spaces
\begin{equation}%41
u = \pm \frac{R}{c}, \quad \hbox{or, what is the same,} \quad v \mp \omega = 0.
\end{equation}
Such spaces may be briefly called tangential spaces.

6. Delimit any two tangential spaces, with a projective aspect,
consider a coherent piece of the world, inside which the hyperboloid does not
penetrate and that according to its shape will in accord with the projective conception
suitably be called a \emph{double wedge}.  This double wedge sticks out
from two sides to the still two-dimensional area, which
is common to the two tangent spaces and that one should therefore conveniently call the \emph{double cutting edge} (of the wedge).

7. The simplest way of getting an overview of this situation is to use the two tangential spaces
considered of No. 5 (in which one can by virtue of the
$G_{10}$ of our collineations every other pair of tangent spaces can conceive in $\infty^4$ ways). The double wedge then covers the points for which
\begin{equation}%42
-\frac{R}{c} < u < +\frac{R}{c}, \quad \hbox{i.e.} \quad -1 < \frac{\upsilon}{\omega} < +1.
\end{equation}
The cutting edge is formed by those points for which $u$ becomes undetermined, thus for which $\upsilon$ and $\omega$ vanish at the same time (for which $x,\ y,\ z$ become infinite).

8. According to the doctrine of projective measurement for everyone of such double wedges, accounting for any two elliptical spaces containing their cutting edge, introduce a real \emph{pseudo-angle}.

9. For the sake of clarity I will follow the example (41), (42). Two associated (with their whole course related to the double wedge) elliptical spaces are then given by the equations:
\begin{equation}
u = u_1, \quad u = u_2, \quad \hbox{or} \quad \frac{\upsilon}{\omega} = \frac{\upsilon_1}{\omega_1}, \quad \frac{\upsilon}{\omega} = \frac{\upsilon_2}{\omega_2}
\end{equation}
(where $u_1$ and $u_2$ are between $\pm \frac{R}{c}$ and $\frac{\upsilon_1}{\omega_1}$, $\frac{\upsilon_2}{\omega_2}$ are between $\pm 1$). They form with the \emph{flanks} of the double wedge, i.e. the two tangential spaces (42), two mutually inverse double ratios of which we want to pick out this:
\begin{equation}%44
D v = \frac{u_1 + R/c}{u_1 - R/c} \cdot \frac{u_2 - R/c}{u_2 + R/c} = \frac{\upsilon_1 + \omega_1}{\upsilon_1 - \omega_1} \cdot \frac{\upsilon_2 - \omega_2}{\upsilon_2 + \omega_2}.
\end{equation}
\emph{The pseudo-angle of the two elliptical spaces (43) then becomes defined by the logarithm of this double ratio multiplied by some real constant $A$.}

\newpage

\noindent 608 \hfill On the Erlangen program \hfill
\bigskip

10. With regard to the de Sitter developments, we want to take $A =\frac{R}{2c}$ and by the way want to set $u_2 = 0$, i.e.\ take the pseudo-angle starting from $u = 0$. \emph{We then have}, while omitting the index of $u_1$, $v_1$, $\omega_1$, \emph{as a definition formula of the pseudo-angle}:
\begin{equation}%45
\varphi = \frac{R}{2c} \log\frac{R/c + u}{R/c - u} = \frac{R}{2c} \log\frac{\omega + \upsilon}{\omega - \upsilon}
\end{equation}
and see clearly how it grows from $-\infty$  to $+\infty$ when $u$ goes from
$-\frac{R}{c}$ to $+\frac{R}{c}$, i.e. wandering through the whole double wedge.

11. For the points of the cutting edge itself, where $\omega$ and $\upsilon$ at the same time
vanish, $\varphi$ naturally becomes completely indefinite. One has with it,
for the general analytical conception, no other singularity
as at the polar angle $\varphi$ at the zero point of an ordinary plane
(polar-)coordinate system. Only that the two absolute directions,
which underlie the determination of angle (in the sense of projective theory), are imaginary in the usual case, but in (45) are real\footnote{
For the Nos. 8--11, the reader who is farther away from these things may like my old developments in Vol. 4 of the Math. Annalen [see Abh. XVI of this edition] (where the relevant relationships and considerations with all verbosity are described).}.

\bigskip
\begin{center}
\S 9.

\medskip
\textbf{Introduction of matter and time.}
\end{center}
\bigskip

We now think of our $ds^2$ (35) by any four independent, provisional parameters $w$ (for which we might take our $x$, $y$, $z$, $u$):
\begin{equation}%46
ds^2 = \sum g_{\mu\nu} dw^\mu dw^\nu.
\end{equation}
Since we know that this $ds^2$ is measured with a constant Riemannian curvature, we can write down the associated $K_{\mu\nu}$ according to the developments of Herglotz\footnote{Saxon reports of 1916, p. 202.}:
\begin{equation}%47
K_{\mu\nu} = \frac{3 c^2}{R^2} \cdot g_{\mu\nu}.
\end{equation}
So it obeys the Einstein field equations with the $\lambda$-term
\begin{equation}%48
K_{\mu\nu} - \lambda g_{\mu\nu} - \chi T_{\mu\nu} = 0,
\end{equation}
if we set
\begin{equation}%49
\lambda = \frac{3c^2}{R^2} \ \ \hbox{and all}\ \ T_{\mu\nu}=0,
\end{equation}
i.e. {\it do not accept matter at all}. We will also see below, that one necessarily

\newpage

\hfill XXXIII. Integral form of conservation laws, etc. \hfill 609
\bigskip

\noindent is lead to this assumption, if one assumes the presupposition of a uniformly fulfilling world of incoherent matter, with an appropriate introduction of a ``time'' $t$ of ``dormant'' matter. In fact, de Sitter comes to this conclusion, except that he puts it a little differently, as one may check on the spot.

Of course, we take off with this formula (49) of Einstein's original physical intention, which was based on a through uniform distribution of matter across the space to provide an average world view. But we also set down in at least a formal contradiction with another principle of Einstein, according to which there should be no non-zero solution of Equations (48) without accepting matter (see the above-cited Einstein's note of March 1918). This principle is with Einstein originally undoubtedly born out of physical considerations, but in itself it is purely mathematical in nature, so it becomes (of which to me Einstein on occasion himself pointed out in correspondence) refuted by the very existence of our $ds^2$ (46). However, one can notice that the $g_{\mu\nu}$ of this $ds^2$ (one might do the calculation for the $x$, $y$, $z$, $u$) along the fundamental hyperboloid becomes infinite which is considered to be an equivalent to the absence of matter at the nonsingular locations of the world.

Let us now focus on the appropriate introduction of a ``time'' $t$ (which we then choose as $w^{IV}$). The starting point, according to Einstein's view, must be the remark that the world we seek as a \emph{static} system should be understood, i.e. that $ds^2$ should remain unchanged if, while holding $w^I$, $w^{II}$, $w^{III}$, the $w^{IV} = t$ increases by any constant. So it should be the single-membered group:
\begin{equation}%50
\bar w^I = w^I, \quad \bar w^{II} = w^{II}, \quad \bar w^{III} = w^{III}, \quad \bar w^{IV} = w^{IV} + C
\end{equation}
be included in the ten-membered group, through which our $ds^2$ passes into itself. A few geometric conclusions suffice to see that such a unitary group is synonymous with a continuing rotation of our pseudo-elliptic world around a fixed, two-dimensional axis, such that $t$ \emph{therefore} (after a suitable choice of the time unit) \emph{except for an additive constant must match with the pseudo angle of a double wedge as defined by} (45). So if we think of any two under $\upsilon = 0$,  $\omega = 0$ in the past understood tangential spaces of the fundamental hyperboloid and don't care about the additive constant we have:
\begin{equation}%51
t = \frac{R}{2c} \log\frac{\omega + \upsilon}{\omega - \upsilon}.
\end{equation}
\smallskip

{\tiny{Klein, Collected math. Treatises. I.\hfil 39 \hfil}}

\newpage

\noindent 610 \hfill On the Erlangen program \hfill
\bigskip

Now there are $\infty^6$ such pairs of tangent spaces. \emph{We after that, according to (51), have $\infty^6$ ways to introduce a $t$} --- in contrast to the cylinder world, where the $t$ is determined except for an additive constant, and in contrast to the special theory of relativity (the Lorentz group), where the $t$ (always after fixing the time unit and the starting point) contains three arbitrary parameters.

First, let us conclude that (51) is exactly the same as the $ds^2$ on which de Sitter bases his hypothesis $B$. On using spatial polar coordinates de Sitter writes (as long as I use the letters I otherwise use, and as well  take the $ds^2$ with the previously agreed sign):
\begin{equation}%52
-ds^2 = \frac{R^2}{c^2} (d\vartheta^2 + \sin^2\vartheta \cdot d\varphi^2 + \sin^2\vartheta \sin^2\varphi \cdot d\psi^2) - \cos^2\vartheta \cdot dt^2
\end{equation}
and this $ds^2$ arises from the one put forward in (35):
$$ -ds^2 = d\xi^2 + d\eta^2 + d\zeta^2 - d\upsilon^2 + d\omega^2, $$
if I, in compliance with condition (34):
$$ \xi^2 + \eta^2 + \zeta^2 - \upsilon^2 + \omega^2 = \frac{R^2}{c^2} $$
simply set:
\begin{equation}%53
\left\{ \begin{aligned}
    \xi &= \frac{R}{c} \sin\vartheta \cos\varphi,  &\eta &= \frac{R}{c} \sin\vartheta \sin\varphi \cos\psi, \\
    \zeta &= \frac{R}{c}\sin\vartheta \sin\varphi \sin\psi, \quad &\upsilon &= \frac{R}{c} \cos\vartheta \, \frak{Sin}\frac{ct}{R}, \\
    \omega &= \frac{R}{c} \cos\vartheta \, \frak{Cos}\frac{ct}{R}.
    \end{aligned}
  \right.
\end{equation}
Here $\frak{Sin}$ and $\frak{Cos}$ are supposed to be hyperbolic functions in the usual way. Then:
\begin{equation}%54
\frak{Tang}\frac{ct}{R} = \frac{\upsilon}{\omega}
\end{equation}
which in fact coincides with formula (51).

The piece of our pseudo-elliptic world, which according to (53), if $\vartheta,~\varphi,~\psi$ within the usual limits, and $t$ however runs from $-\infty$ to $+\infty$, I am going to call a {\it de Sitter world}.
According to (54) $\frac{\upsilon}{\omega}$ goes through only the values of $- 1$ to $+ 1$. Apparently this de Sitter's world is nothing else than the {\it double wedge} of previous paragraphs.  Its two ``flanks'', $\upsilon - \omega = 0$ and $\upsilon + \omega = 0$, appear as the infinitely distant future, or the infinitely distant past.
Its edge, however (which for the general conception of the pseudoelliptic world consists of all ordinary points) appears as something singular, namely as the place of such world points, for which $t$ has the value $0 /0$.
---

\newpage

\hfill XXXIII. Integral form of conservation laws, etc. \hfill 611
\bigskip

I already have touched these relationships mentioned in the above point at the
Annual Report of the German Mathematical Society (lecture before the G\"ottingen Mathematical Society of June 11, 1918).
To the paradoxical relations, which for the physical conception at
present, clearly stand out as such, I have at that time
uttered: ``Two astronomers who, both living in a de Sitter world, are equipped with various de Sitter watches
would be able to talk in terms of the reality or imaginativeness of any world events in a very interesting way.''
What is meant is that the double wedges, which from the pseudo-elliptical world
through different pairs of tangential spaces of the fundamental
hyperboloids are cut out, always have only pieces in common,
with other pieces extending each other. ---

For the rest, whoever wishes, can easily orient themselves more precisely about the details of de Sitter's world. The world is only approaching enough in the two points: $\xi = 0$, $\eta = 0$, $\zeta = 0$, $\upsilon \mp \omega = 0$ to the fundamental hyperoloid.
All world lines are such conic sections which touch the hyperboloid in these two points (whose plane thus contains the one-dimensional axis $\xi = 0$, $\eta = 0$, $\zeta = 0$).
There is only one continuous $G_4$ left, which transforms the de Sitter world into itself, according to the substitution $\bar t = t + C$ associated with the continuous $G_3$ of the unimodular orthogonal substitutions of $\xi$, $\eta$, $\zeta$. Here $\xi^2 + \eta^2 + \zeta^2$ is invariant, so the group of de Sitter's world is no longer transitive. The ``axis'' $\xi = 0$, $\eta = 0$, $\zeta = 0$ and
the ``cutting edge'' $\upsilon = 0$, $\omega = 0$ are invariant entities.

Finally, we convince ourselves that the density $\varrho$ of the resting, incoherent matter, which is to uniformly fulfill the de Sitter world, in fact necessarily $= 0$ must be set. Let us stay with our ``static'' coordinates. We have then for all other index combinations $\mu,~\nu$:
$$ \lambda g_{\mu\nu} = \frac{3 c^2}{R^2} g_{\mu\nu} $$
and only for $\mu = 4$, $\nu = 4$:
$$ \lambda g_{44} = \frac{3 c^2}{R^2} g_{44} + \chi c^2 \varrho, $$
from which clearly follows
$$ \lambda = \frac{3 c^2}{R^2}, \qquad \varrho = 0 $$
as we had already assumed in formula (49). ---

All of these results are in full agreement with de Sitter's own statements. But they contradict the objection, that Einstein raised in his communication of March

\hfil {\tiny{39*}} \qquad

\newpage

\noindent 612 \hfill On the Erlangen program \hfill
\bigskip

\noindent 1918 against de Sitter and then Weyl in his book\footnote{
Space, time, matter. P. 225.},
and recently in a special essay in the Physical Journal\footnote{
1919, No. II (dated January 15, 1919).}
has supported by detailed calculations. Both authors find that along the cutting edge of the double wedge (For brevity I will stay in my idiom) matter must be existing. I have not had the correctness of Weyl's calculations verified, however, I would like to agree with Einstein's opinion by letter pronounced that the difference of the mutual results must be justified by the diversity of the coordinates used.
What I, using the $\xi$, $\eta$, $\zeta$, $\upsilon$, $\omega$ name as a single point on the cutting edge is, when using the $\vartheta$, $\varphi$, $\psi$, $t$ (because of the indefinite value of $t$) a simply extended area.

My final vote however on the de Sitter statements is that mathematically --- at least except for this one not yet completely clarified point [which I would like to see explained in a general way] --- everything is fine, but one is led to physical conclusions, which contradict to our ordinary way of thinking and at least to the intentions, which Einstein pursued when he introduced the spatially closed world.

\end{document}